\documentclass{emulateapj}
\usepackage{psfig}
\usepackage{apjfonts,times}

\newcommand{\beq}{\begin{equation}}
\newcommand{\eeq}{\end{equation}}

\newcommand{\lapprox}{$\stackrel {<}{_{\sim}}$}

\def\farcm{\hbox{$.\mkern-4mu^\prime$}}

\def\earth{\hbox{$\oplus$}}
\def\arcmin{\hbox{$^\prime$}}

\newcommand{\AmS}{{\protect\the\textfont2
  A\kern-.1667em\lower.5ex\hbox{M}\kern-.125emS}}
\newcommand{\lsim}{\ \raise
-2.truept\hbox{\rlap{\hbox{$\sim$}}\raise5.truept\hbox{$<$}\ }}
\newcommand{\gsim}{\ \raise
-2.truept\hbox{\rlap{\hbox{$\sim$}}\raise5.truept\hbox{$>$}\ }}
\newcommand{\simsim}{\ \raise
-2.truept\hbox{\rlap{\hbox{$\sim$}}\raise5.truept\hbox{$\sim$}\ }}

\slugcomment{Astrophys. J. Letters 652 (2006) L93}

\shorttitle{HST/WFPC2 Photometry of NGC 2011} 
\shortauthors{Gouliermis et al.}

\begin{document}

\title{HST WFPC2 Observations of the Peculiar Main Sequence of the Double
Star Cluster NGC 2011 in the Large Magellanic Cloud}

\author{D. A. Gouliermis\altaffilmark{1}, 
        S. Lianou\altaffilmark{2}, 
        M. Kontizas\altaffilmark{2}, 
        E. Kontizas\altaffilmark{3},
        A. Dapergolas\altaffilmark{3}}
\altaffiltext{1}{Max-Planck-Institut f\"{u}r Astronomie, K\"{o}nigstuhl 
17, 69117 Heidelberg, Germany, dgoulier@mpia.de}
\altaffiltext{2}{National and Kapodistrian University of Athens, Faculty 
of Physics, Dpt of Astrophysics, Astronomy and Mechanics, 
Panepistimiopolis, Zografos, 15784 Athens, Greece, 
s\_lianou@phys.uoa.gr, mkontiza@phys.uoa.gr}
\altaffiltext{3}{National Observatory of Athens, Institute of Astronomy 
and Astrophysics, Lofos Nymfon, Thiseio, P.O.Box 20048, 11810 Athens, 
Greece, ekonti@astro.noa.gr, adaperg@astro.noa.gr}

\begin{abstract}

We report the serendipitous discovery of a peculiar main sequence in 
archived Hubble Space Telescope WFPC2 observations of the young star 
cluster NGC 2011 in the Large Magellanic Cloud. The bright part of this 
main sequence exhibits a prominent double, fork-like feature, as if it 
consists of twin main sequences, one of them being redder. The 
color-magnitude diagram, constructed from the stars found in the only 
available WFPC2 field of the cluster, is used to distinguish the stars 
according to their membership to each of these sequences and to study 
their spatial distribution. We find that there are two well distinguished 
populations in the sense that the redder main sequence is dominated by 
stars that belong to the main body of the cluster, while the stars of the 
bluer main sequence belong to the surrounding region. Providing that NGC 
2011 is a verified binary cluster, with the second companion unfortunately 
not observed, and taking into account the general region where this 
cluster is located, we discuss the possible scenarios from both star 
formation, and early dynamical evolution point-of-view that might explain 
this unique discovery.

\end{abstract}


\begin{figure*}[t!]
\centerline{\hbox{
\psfig{figure=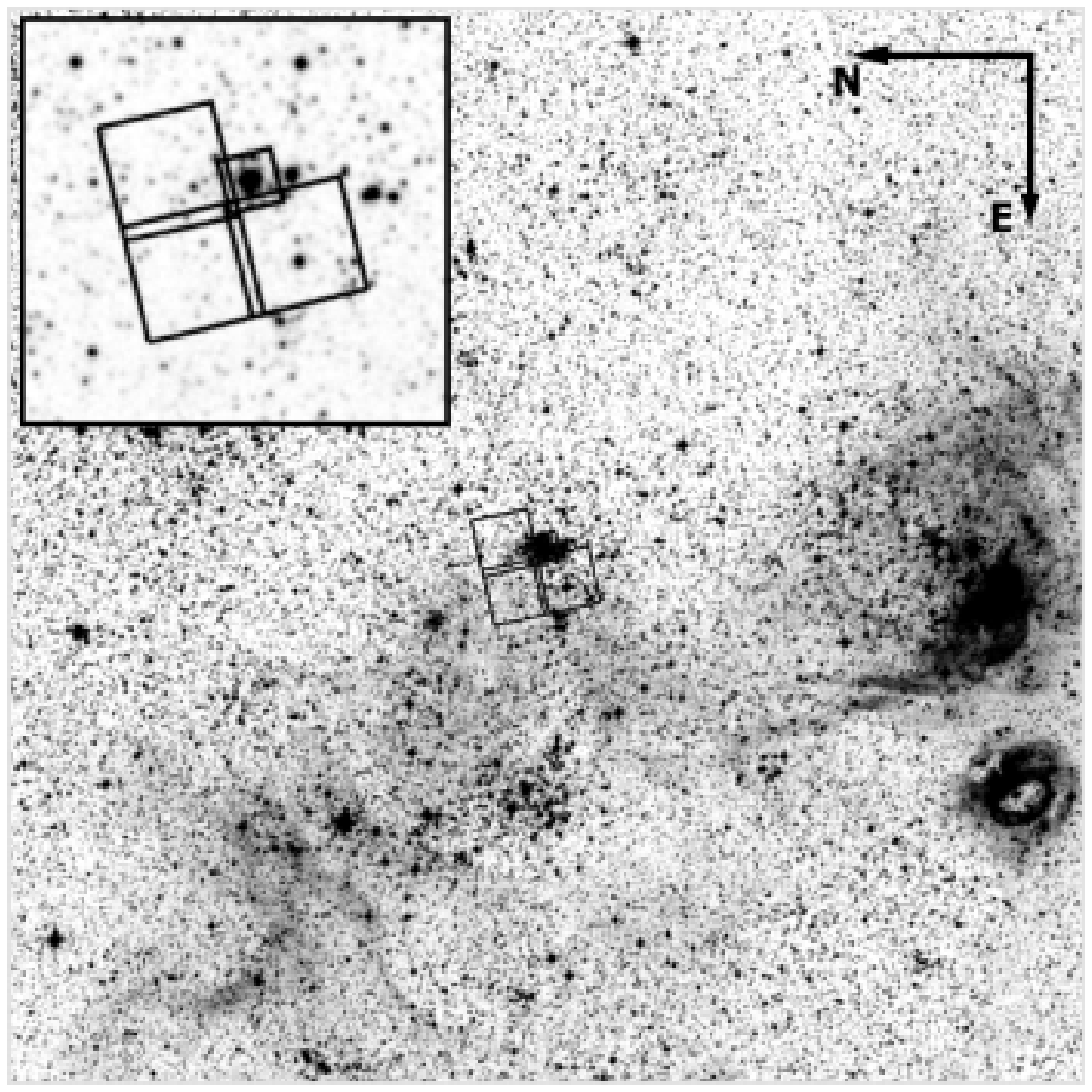,width=0.4350\textwidth,angle=0}
\psfig{figure=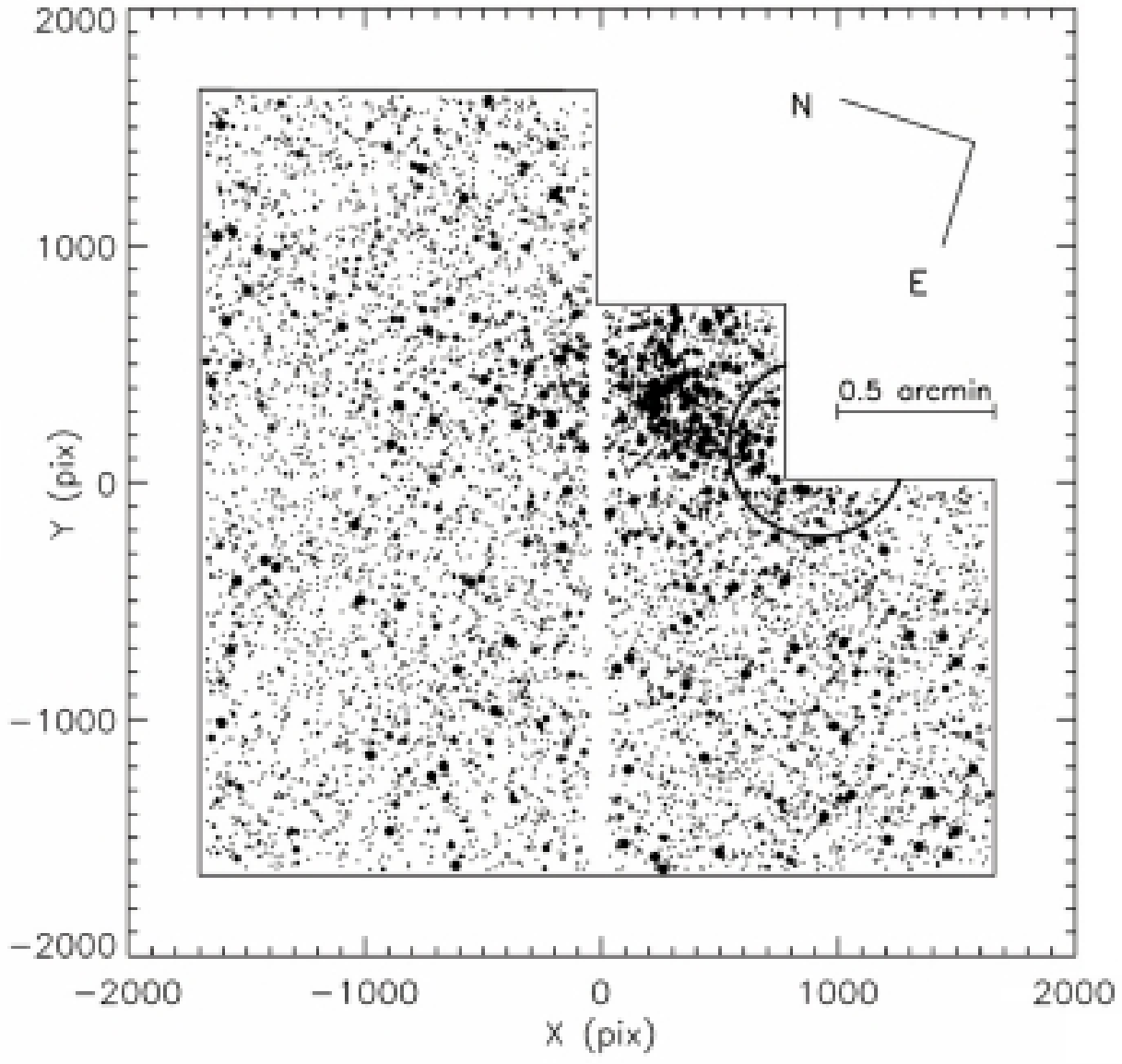,width=0.4550\textwidth,angle=0}
}}
\caption{Left: A 25\arcmin\ $\times$ 25\arcmin\ image of the
general region, where NGC 2011 is located with the observed WFPC2 field
overplotted. This image is selected from a digitized wide-field UKST
photographic plate in the B-band, to exhibit the nebulosity of the
region. The observed WFPC2 field-of-view (FoV) was such that the PC
frame covers the primary cluster NGC 2011a. The second companion is
unfortunately outside this FoV (located to the right of PC frame) as it
is shown in the inset image, constructed from an I-band UKST
photographic plate, which does not show any emission from the nebula.
Right: Map of stars found with HSTphot photometry based on
WFPC2 imaging of NGC 2011. An outline is drawn to indicate the position
of the possible component NGC 2011c. Component NGC 2011a is almost
entirely covered by the PC frame of WFPC2. The images in both panels
have almost identical orientation for reasons of comparison.}
\label{fig-map} 
\end{figure*}

\section{Introduction}

The Large Magellanic Cloud (LMC) contains an extraordinary sample of star 
clusters with a large variety in spatial distribution, age and luminosity 
(e.g. Kontizas et al. 1990; Bica et al. 1999). This intriguing population 
of clusters include peculiar types of systems, like binary and elliptical 
clusters, that we do not see in the Galaxy. In this letter we present our 
findings from {\em Hubble Space Telescope} (HST) observations of the LMC 
cluster NGC 2011 with the {\em Wide Field Planetary Camera 2} (WFPC2). NGC 
2011 is a young star cluster with age 6 ($\pm$ 1) Myr (Elson \& Fall 1988) 
located in a young region (Figure \ref{fig-map} left) at the southern edge 
of the super-giant shell LMC 4 (Meaburn 1980). It is considered as one of 
the brightest and most populous double clusters in the LMC (Kontizas et 
al. 1989). It was originally identified as a probable double cluster by 
Bhatia \& Hatzidimitriou (1988) who conclude that among the LMC cluster 
population binary clusters constitute a statistically significant sample.

The multiplicity of the cluster, in the sense that projection effects 
cannot account for this pair, was verified by Kontizas et al. (1993), who 
studied the stellar content in the outer region of the cluster by means of 
spectra classification from UK Schmidt objective prism spectra, and in the 
core of the cluster with low resolution {\em International Ultraviolet 
Explorer} (IUE) spectra. From the spectral classification it was found 
that both members of the pair show identical stellar content, dominated by 
early type stars, quite different from the stellar component of the 
neighboring field, where stars with spectral type later than A dominate. 
The UV integrated spectra taken with IUE showed that the stellar content 
in the core of both the brighter (named NGC 2011a) and fainter (NGC 2011b) 
members of the pair are found to be very similar. Kontizas et al. (1993) 
note that the core of NGC 2011a is elongated, so that another UV feature, 
which was named NGC 2011c, at an off-center position, but still in the 
dense core to be observed. The IUE spectra of NGC 2011b and NGC 2011c are 
found to be very similar.


Unfortunately the second component, NGC 2011b, is not covered in the WFPC2 
field presented here, but the elongation of the core of NGC 2011a is very 
well shown in these observations, and a part (if not all) of the third 
possible component, NGC 2011c, can be seen at the southern corner of the 
PC frame of the camera (Figure \ref{fig-map} Right). In this Letter we 
report the peculiar behavior of the main sequence of NGC 2011, and we 
discuss its relation to the multiple nature of the cluster.

\section{ Observations and Photometry}

The results presented here are based on a single WFPC2 field taken with 
the PC frame centered on NGC 2011a. The data were taken in filters F555W 
($\sim V$) and F814W ($\sim I$) as part of the program GO-8134, and we 
retrieved them from the HST data archive\footnote{available from STScI at 
{\tt http://archive.stsci.edu/hst/search.php} and ESO at {\tt 
http://archive.eso.org/wdb/wdb/hst/science/form}}. The six datasets 
(archive names U5AY0801R - U5AY0806R) include three exposures (2 $\times$ 
350 sec and 1 $\times$ 10 sec) for each filter. The data were reduced 
using the HSTphot photometry package (Dolphin 2000). A detailed account of 
the photometry process with HSTphot is given by Gouliermis et al. (2005).

We combined the two images of 350 seconds in each filter using the
subroutine {\it{coadd}} to produce deep exposures. HSTphot is suited for
HST/WFPC2 observations and it is especially designed to perform
simultaneous photometry on both short- and long-exposure images and for
more than one filters. We, thus, performed our photometry on both deep
(700 sec) and shallow (10 sec) images observed in both filters to
produce the final photometric catalog of 6,760 stars found in the
observed field.  The detection limit of the short exposures is $V
\approx$ 21.5 mag, and the brightest magnitude observed in the long
exposures is $V \approx$ 18.5 mag. The present study is part of an
ongoing investigation of the phenomenon of mass segregation in peculiar
LMC clusters, and more details on the photometry will be given in a
forthcoming paper (S. Lianou et al., in preparation).

\begin{figure}
\epsscale{1.}
\plotone{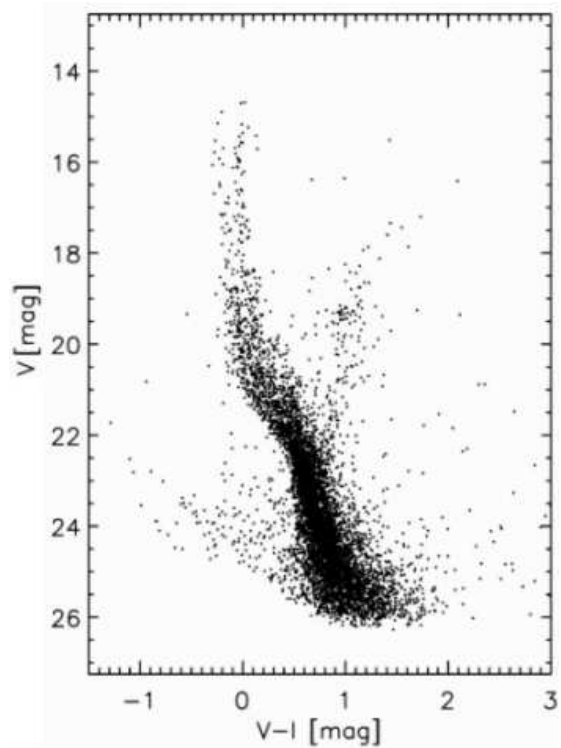}
\caption{$V-I$, $V$ Color-Magnitude Diagram (CMD) of all stars detected in 
the area of NGC 2011 with WFPC2 imaging. The bright fork-like feature of 
the main sequence, that seems to consist of two distinct sequences, is 
quite prominent. One of the branches of this feature is located to the red 
side of the other and both show to be populated by the same kind of stars. 
Are these two bright main sequences indicative of different stellar 
systems, or are they part of the same peculiar stellar concentration?}
\label{fig-cmd} 
\end{figure}

\begin{figure}
\epsscale{1.}
\plotone{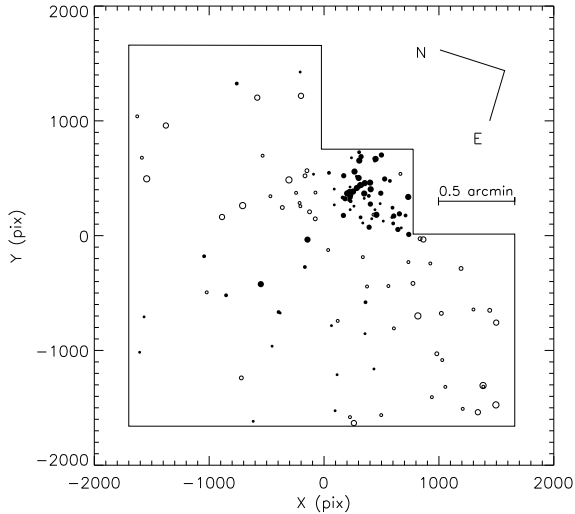}
\caption{Map of the bright stars of the Blue (open circles) and the Red
(filled circles) main sequence. This map, which codes stellar brightness
by symbol size, exhibits the discrepancy of the loci of the stars
according to their membership to one of the branches of the fork-like
main sequence observed in the area of NGC 2011. Most of the Red MS stars
are strictly confined within the cluster itself (within $\sim$ 0\farcm4
from its geometrical center), while the stars of the Blue MS are
covering the remaining surrounding area, and they avoid the cluster
region.}
\label{fig-mapsel} 
\end{figure}

\section{Stellar Populations in the Observed Field}

\subsection{The Fork-like Main Sequence}

The color-magnitude diagram (CMD) of the stars found in the observed area 
of NGC 2011 is shown in Figure \ref{fig-cmd}. In this CMD the upper main 
sequence shows to have two distinct branches, one redder than the other, 
that form a fork-like pattern. This feature is observed in the stellar 
samples from both short and long exposures. The obvious resemblance of 
these two main sequences suggest that they are populated by similar kinds 
of stars and that the shift of the one to the red is probably due to 
reddening. Indeed, as it will be shown later the right sequence (from 
hereon we will refer to it as the ``Red'' main sequence) seems to fit the 
same isochrone models as the left (``Blue'' main sequence) but assuming 
higher extinction.

In order to distinguish the areas where the stars of each of these
sequences are located and to check for any difference in their spatial
distributions, we selected two regions in the CMD, covering the bright
stars that populate each of the sequences, and we constructed the
corresponding stellar maps. We selected stars down to $V = 19$ mag,
which is the faintest magnitude where the two sequences are
discriminable. The map of both stellar groups is shown in Figure
\ref{fig-mapsel}, where the stars of the Blue main sequence (MS) are
plotted with open and of the Red MS with filled circles. In this map the
loci of the bright stars representative of each sequence can be easily
distinguished. Specifically, the Red MS is found to correspond to the
main body of NGC 2011 (within $\sim$ 0\farcm4 from its center), observed
in the PC frame of the WFPC2 (in addition to a loose strip of stars
outwards to the east/north-east), while the stars of the Blue MS are
located in the surrounding region, and they ``avoid'' the main cluster.
This observation indicates that there is preferably higher extinction in
the area of the cluster than in the surrounding region.

Indeed, as shown in Figure \ref{fig-map}, NGC 2011 is located in a region 
of high nebulosity. This region belongs to a stellar aggregate, along with 
several stellar associations and other young clusters in the stellar 
super-complex SC2. This complex, which is classified as an ``active star 
forming complex'' (Livanou et al. 2006) is located at the southern edge of 
the super-giant shell LMC 4, the borders of which are characterized by 
recent star formation (e.g. Yamaguchi et al. 2001). An examination of the 
region of NGC 2011 in the Magellanic Cloud Emission Line Survey (MCELS; 
Smith et al. 2005) showed no trace of interstellar gas that has been 
heated and energized by stars. On the other hand, observations with 
Spitzer 
within the SAGE survey (Meixner et al. 2005) show a prominent bright 
infrared source on the cluster itself, giving clear indications of recent 
star formation in the region of NGC 2011. Could it be that NGC 2011 is a 
cluster on the act of forming stars? The observations presented so far 
seem to support this idea.


\subsection{Characterization of the Observed Stellar Populations}

If we assume that each of the two branches of the bright fork-like MS 
(Blue and Red) represents a stellar sample with a fully populated mass 
function (down to the detection limit) it would be interesting to see the 
whole CMD of each of these populations. Therefore, considering that each 
one of the areas, which covers stars of the Blue and Red main sequences is 
well defined, the {\em faint} part of the corresponding CMD can easily be 
found if we select stars confined only in each of these areas. The PC 
frame is found to cover almost all of the brightest stars of the Red MS, 
without any contamination by stars of the Blue MS and therefore we 
selected all stars found in the PC frame, as well as in the small strip 
outwards from the cluster to the east/north-east direction (Figure 
\ref{fig-mapsel}) to construct the corresponding CMD down to the detected 
faintest magnitudes. For the full magnitude range CMD, which corresponds 
to the Blue MS, we selected two boxed areas, to the north-east and south 
of the PC frame, which are found to cover most of the bright young 
stars of the Blue MS (Figure \ref{fig-mapsel}).

\begin{figure*}
\plotone{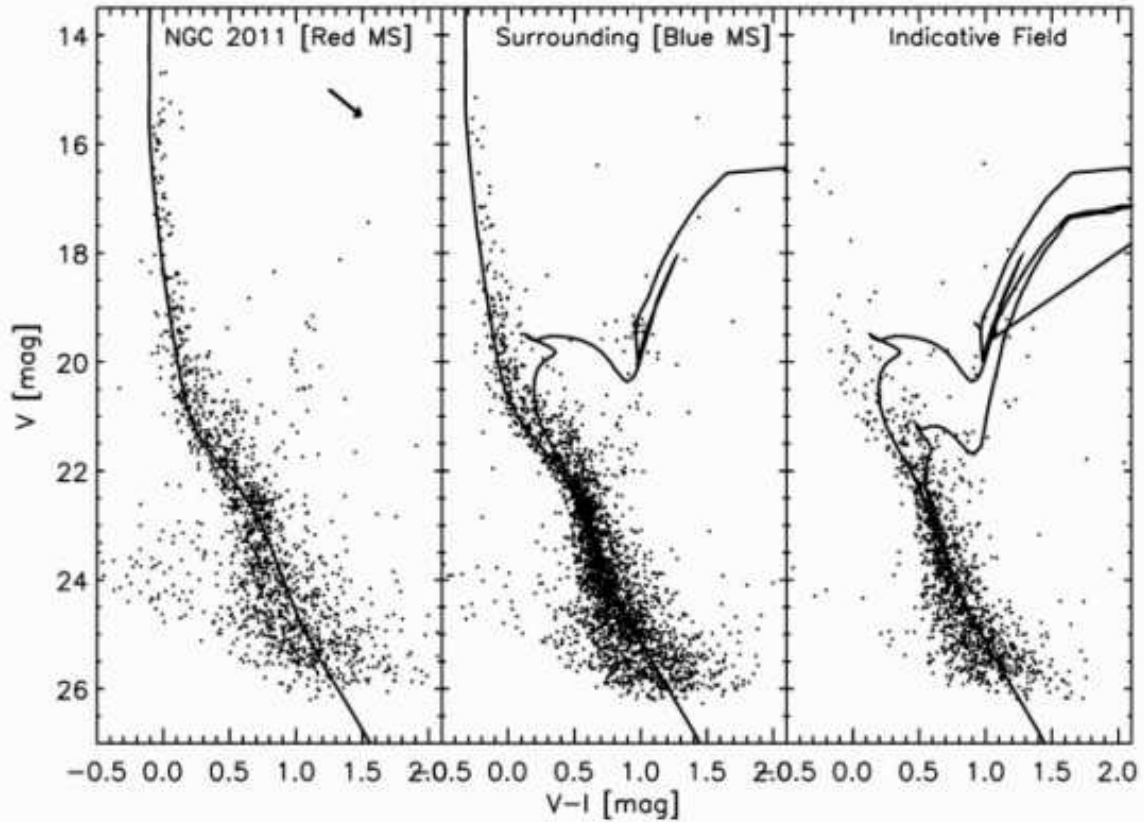}
\caption{The $V-I$, $V$ CMDs of the stars from our catalog confined
within (1) the PC frame of WFPC2 and a strip east/north-east from it,
which covers the Red MS in the main body of NGC 2011 (left panel), (2)
two boxed areas, one to the north-east and one to the south of the first
(Figure \ref{fig-mapsel}), which cover the population of the Blue MS
(center panel), and (3) to the most distant east and north corners of
the FoV, which provide an indicative population of the LMC general field
(right panel). Isochrone models are overlayed. The model for
$\log{\tau}=6.6$ is overplotted on the Red MS with reddening $A_{\rm V}
\simeq$ 0.48 mag applied (the corresponding vector is shown on the right
of the MS), and on the Blue MS with a modest reddening of $A_{\rm V}
\simeq$ 0.03 mag. In the central and right panels isochrones
representative of the field population of the LMC for ages 1 Gyr and 1
and 3 Gyr respectively are also shown. These isochrone fits show that
stars of both Blue and Red MSs seem to have the same age with
$\log{\tau}\leq 7$, but the rddening in the corresponding areas is quite
different.}
\label{fig-cmdsys}
\end{figure*}

Considering that the WFPC2 FoV is too small to also cover a useful part of 
the general field of the LMC, and that the observed region is not 
representative of this field (Figure \ref{fig-map}), one cannot expect to 
have any good measurement of the contribution of the LMC field population 
to the observed CMDs. Nevertheless, we selected the two most distant 
corners of the observed FoV (one to the far east and one to the far north 
of the PC frame) for plotting an indicative CMD of the general LMC field. 
The CMDs of all three selected areas are shown in Figure \ref{fig-cmdsys}. 
It can be seen that indeed the whole magnitude range of stars, which 
belong to the Red and Blue MSs can be distinguished. This was further 
verified by applying isochrone fitting to each of the CMDs shown in Figure 
\ref{fig-cmdsys}. The evolutionary models of the Padova group in the HST 
WFPC2 VEGA magnitude system (Girardi et al. 2002) were used.

We found that both Blue and Red MSs are young with an age not older than 
about $10^{7}$ yr in agreement with the result of Elson \& Fall (1988). 
Still there is a strong difference in reddening between the Blue and Red 
MSs. Specifically, we found that the CMD of the Blue MS shows a modest 
color excess of $E(B-V) \simeq 0.01$ mag, equal to the one found for the 
field, while the color excess found from model fitting on the Red MS CMD 
is almost an order of a magnitude higher, $E(B-V) \simeq 0.15$ mag, which 
corresponds to optical extinction $A_{V} \simeq 0.48$, assuming the 
reddening law $R_V = 3.2$ (Mihalas \& Binney 1981). This result verifies 
the suggestion mentioned earlier, that although both MSs are populated by 
similar kind of stars, there is a clear discrepancy in the absorption in 
each of the corresponding area, indicating that the main body of NGC 2011 
is embedded, strongly suggesting a recently formed star cluster, which did 
not have the time to expel its gas. Still, it should be noted that the 
suggested reddening for NGC 2011 is much higher than the one given by van 
den Bergh (1981) who presented integrated colors for 147 LMC clusters.

\section{Final Remarks}

We showed evidence of the coexistence of two stellar groups in the same
region of the LMC cluster NGC 2011, as it is observed with WFPC2. This
is exhibited by a double pattern in the bright end of the observed main
sequence. We found that each branch of this fork-like feature represents
a fully populated main sequence (down to the detection limit), showing
indications of twin stellar populations, the one located in the cluster
itself being still embedded in its star forming gas. If these two
populations belong to the same system or not is open for discussions,
and the results presented so far rise interesting questions concerning
the relation between the peculiar main sequence of NGC 2011 and its
binary nature. Considering that the data presented here are the best
available today of this cluster, with no kinematic information on
individual stars in the region, we can only speculate on the scenarios
that explain our observations.

The simplest explanation would be that both main sequences belong to the 
cluster itself, but some of the stars are not obscured by nebulosity, 
which is concentrated in the core of NGC 2011. But, in order to explain 
the existence of massive stars away from the core, since dynamical 
evolution would tend to segregate these stars to the center (Meylan \& 
Heggie 1997), one should assume that the cluster is under disruption. 
Indeed, the elongated core observed in our WFPC2 field supports this idea. 
Furthermore, the surface density profiles constructed for NGC 2011 within 
our study of mass segregation (S. Lianou et al., in preparation) 
deviate from King (1962) profiles, indicative of tidal stripping, although 
King models may not be the most suitable for such clusters, which are not 
tidally truncated (Elson et al. 1987). Still, NGC 2011 is a binary cluster 
with indications of tidal interaction from the companion cluster NGC 2011b 
(Kontizas et al. 1993), and with a third component observed in our WFPC2 
data, as a stellar ``bridge'' between NGC 2011a and NGC 2011b (Figure 
\ref{fig-map}).

Under these circumstances, and since the stellar populations of both 
members of the binary cluster are identical, one may ask if the stars in 
the Blue MS are not members of NGC 2011a, but of the second companion, NGC 
2011b, stripped away from their host through dynamical interaction with 
the primary. According to recent N-body simulations, clusters with an 
initial separation smaller than 1\arcmin - 1\farcm3 tend to merge in 
\lapprox\ 60 Myr due to loss of angular momentum from escaping stars 
(Portegies Zwart \& Rusli 2006). Considering that the actual center to 
center separation of the NGC 2011 pair is 0\farcm95 (Kontizas et al. 1993) 
this is a possible future for this system. Are the two members of NGC 2011 
binary cluster very young clusters in the process of early dynamical 
merging? It is probable, but more observations preferably with the wider 
FoV of WFC of the {\em Advanced camera for Surveys} on HST will definitely 
give us a more clear-cut answer.

\acknowledgments

D. Gouliermis acknowledges the support of the German Research Foundation 
(Deutsche Forschungsgemeinschaft - DFG) through the individual grant 
1659/1-1. M. Kontizas and E. Kontizas wish to thank the General 
Secretariat for Research and Technology and PYTHAGORAS II, project funded 
by the European Social Fund and National Resources (EPEAEK II), for 
financial support. This paper is based on observations made with the 
NASA/ESA Hubble Space Telescope, obtained from the data archive at the 
Space Telescope Science Institute. STScI is operated by the Association of 
Universities for Research in Astronomy, Inc. under NASA contract NAS 
5-26555.


\end{document}